\documentclass[twocolumn,english,aps,prb,superscriptaddress]{revtex4}
\usepackage{graphicx}
\usepackage{multirow}

\begin{document}

\title{Magnetic excitations in underdoped Ba(Fe$_{1-x}$Co$_{x}$)$_{2}$As$_{2}$ with $x=$0.047}

\author{G.~S.~Tucker}
\affiliation{Ames Laboratory and Department of Physics and Astronomy, Iowa State University, Ames, IA 50011, USA}
\author{R.~M.~Fernandes}
\affiliation{Department of Physics, Columbia University, New York, NY 10027, USA }
\affiliation{Theoretical Division, Los Alamos National Laboratory, Los Alamos, NM 87545, USA}
\author{H.-F.~Li}
\affiliation{Ames Laboratory and Department of Physics and Astronomy, Iowa State University, Ames, IA 50011, USA}
\author{V.~Thampy}
\affiliation{Department of Physics and Astronomy, Johns Hopkins University, Baltimore, MD 21218, USA}
\author{N.~Ni}
\affiliation{Ames Laboratory and Department of Physics and Astronomy, Iowa State University, Ames, IA 50011, USA}
\author{D.~L.~Abernathy}
\affiliation{Quantum Condensed Matter Division, Neutron Sciences Directorate, Oak Ridge National Laboratory, Oak Ridge, TN 37831, USA}
\author{S.~L.~Bud'ko}
\affiliation{Ames Laboratory and Department of Physics and Astronomy, Iowa State University, Ames, IA 50011, USA}
\author{P.~C.~Canfield}
\affiliation{Ames Laboratory and Department of Physics and Astronomy, Iowa State University, Ames, IA 50011, USA}
\author{D.~Vaknin}
\affiliation{Ames Laboratory and Department of Physics and Astronomy, Iowa State University, Ames, IA 50011, USA}
\author{J. Schmalian}
\affiliation{Institut f\"ur Theorie der Kondensierten Materie, Karlsruhe Institut f\"ur Technologie, D-76131 Karlsruhe, Germany}
\author{R.~J.~McQueeney}
\affiliation{Ames Laboratory and Department of Physics and Astronomy, Iowa State University, Ames, IA 50011, USA}

\begin{abstract}
The magnetic excitations in the paramagnetic-tetragonal phase of underdoped Ba(Fe$_{0.953}$Co$_{0.047}$)$_2$As$_2$, as measured by inelastic neutron scattering, can be well described by a phenomenological model with purely diffusive spin dynamics. 
At low energies, the spectrum around the magnetic ordering vector $\mathbf{Q}_{\mathrm{AFM}}$ consists of a single peak with elliptical shape in momentum space.
At high energies, this inelastic peak is split into two peaks across the direction perpendicular to $\mathbf{Q}_{\mathrm{AFM}}$. 
We use our fittings to argue that such a splitting is not due to incommensurability or propagating spin-wave excitations, but is rather a consequence of the anisotropies in the Landau damping and in the magnetic correlation length, both of which are allowed by the tetragonal symmetry of the system. 
We also measure the magnetic spectrum deep inside the magnetically-ordered phase, and find that it is remarkably similar to the spectrum of the paramagnetic phase, revealing the strongly overdamped character of the magnetic excitations.
\end{abstract}
\maketitle

\section{Introduction}

The discovery of high-temperature superconductivity in iron-based compounds \cite{Kamihara08,Rotter08} triggered an intense research activity to understand these materials \cite{Canfield10,Johnston10}. 
By now, there is ample evidence indicating the unconventional nature of the superconducting state, such as the existence of a magnetic resonance mode\cite{Christianson08}; the observation of half-integer flux-quantum in niobium-iron pnictide loops \cite{Chen10}; the microscopic coexistence between magnetism and superconductivity \cite{Fernandes10}; and the linear temperature-dependence of the low-$T$ superfluid density in some compounds \cite{Hashimoto10}. 
Due to the proximity of the superconducting transition temperature ($T_\mathrm{c}$) to a magnetic instability, one of the main candidates to explain this unconventional state is the pairing mechanism mediated by spin fluctuations \cite{Mazin09}.

In order to investigate the viability of spin fluctuations as the pairing mechanism, it is paramount to have a consistent and complete description of the magnetic excitations. 
Inelastic neutron scattering (INS) measurements performed by several groups have provided a common picture for the so-called 122 compounds (\emph{A}Fe$_{2}$As$_{2}$, \emph{A} denoting alkaline earth)\cite{Lester10,Diallo10,Li10,Park10,Harriger11,Ewings11}.
Similar to the cuprates, the iron pnictide superconductors are characterized by high magnetic energy scales extending beyond 200 meV. 
Such high energy scales could possibly support high temperature superconductivity where electrons are paired by spin fluctuations.\cite{Letacon11} In the iron pnictides, the striped magnetic ground state is characterized by the ordering vector $\mathbf{Q}_{\mathrm{AFM}}=(\frac{1}{2},\frac{1}{2},1)$ (in tetragonal notation), with antiferromagnetic spin correlations in the direction parallel to $(\frac{1}{2},\frac{1}{2},0)$ and ferromagnetic spin correlations along the direction perpendicular to $(\frac{1}{2},\frac{1}{2},0)$. 

Unlike the N\'eel (or checkerboard) AFM ordered state, this striped AFM state breaks the tetragonal symmetry of the lattice, and results in an orthorhombic distortion with the longer (antiferromagnetic) axis $a$ parallel to $\mathbf{Q}_{\mathrm{AFM}}$ and the shorter (ferromagnetic) axis $b$ perpendicular to $\mathbf{Q}_{\mathrm{AFM}}$.  
However, in several systems, the tetragonal symmetry-breaking, probed by the development of the orthorhombic distortion, takes place above the magnetic transition temperature $T_\mathrm{N}$, implying that long-range magnetic order is not the origin of the structural transition.
Rather, it is possible that magnetic fluctuations can spontaneously break the tetragonal symmetry in the paramagnetic (PM) phase, giving rise to the so-called electronic nematic state.\cite{Fang08,Xu08} 
Thus, the systematic study of the magnetic spectrum via INS is important not only to shed light on the superconducting pairing mechanism, but also to clarify the nature of the tetragonal symmetry-breaking in the iron pnictides.

INS studies have focused on the magnetic excitation spectrum in both the parent compounds, which are antiferromagnetic (AFM) metals, and doped superconducting (SC) compositions that are characterized by strong paramagnetic spin fluctuations. 
In each case, the INS measurements of the spin excitations have revealed surprisingly strong in-plane anisotropies. 
For instance, the orthorhombic-AFM state of the parent CaFe$_{2}$As$_{2}$ compound displays a significant in-plane anisotropy in the spin-wave velocities and in the zone-boundary energies \cite{Diallo09,Zhao09}.
When the spin wave spectrum is fitted with the $J_{1}-J_{2}$ Heisenberg model, the only way to account for such anisotropies is to use very different values for the nearest-neighbor exchange interactions in the directions parallel ($J_{1a}$) and perpendicular ($J_{1b}$) to $\mathbf{Q}_\mathrm{AFM}$ - to the extent that one is negative (antiferromagnetic) and the other positive (ferromagnetic).\cite{Zhao09} 
While $J_{1a}\neq J_{1b}$ is justified in the orthorhombic-AFM phase, the disparity between their values, especially considering the rather small orthorhombic distortion $a-b$, suggests that a simple description in terms of localized moments may not suffice. 
Even in the tetragonal-PM phase of CaFe$_{2}$As$_{2}$, where the low-energy spin fluctuations were shown to be diffusive in nature, the magnetic correlation lengths perpendicular and parallel to $\mathbf{Q}_\mathrm{AFM}$ are unequal. 
Such an in-plane anisotropy does not contradict the tetragonal symmetry of the high-temperature phase as $\mathbf{Q}_{\mathrm{AFM}}=(\frac{1}{2},\frac{1}{2},1)$ is a point of two-fold symmetry located on the Brillouin zone boundary. 
Similar anisotropic spin fluctuations have also been observed in the tetragonal phase of the Ba(Fe$_{1-x}$Co$_{x}$)$_{2}$As$_{2}$ samples with $x=0.065$ \cite{Ewings11} and $0.074$ \cite{Li10}, where long-range AFM order is absent. 
Whether the observed anisotropies of the spin-wave velocities in the orthorhombic-AFM phase and of the magnetic correlation lengths in the tetragonal-PM phase share a common origin remains an open issue.

At high energies ($> 80$\,meV), anisotropic spin fluctuations within the Fe layer develop a transverse splitting in paramagnetic Ba(Fe$_{1-x}$Co$_{x}$)$_{2}$As$_{2}$ with $x =$ 0.074.\cite{Li10}
This unusual spectral feature has also been recently observed in the parent BaFe$_{2}$As$_{2}$ compound.\cite{Harriger11} 
Although different groups generally agree on these observations, the nature of the high-energy peak-splitting is currently under intense debate. The work by Li \emph{et al.}, which first observed such splitting, attributed it to the existence of quasi-propagating modes with different velocities along the transverse and longitudinal directions \cite{Li10}, without, however, providing an explanation for their highly anisotropic character. 
Based on band-structure calculations, Park \emph{et al.}~associate the high-energy splitting to an incommensuration of the SDW ordering vector \cite{Park10}, whose detection apparently falls below the resolution limit in the low-energy regime. 
Harriger \emph{et al.} performed a detailed study of the high-energy spectrum in both the orthorhombic and tetragonal phases of the parent compound $\mathrm{BaFe_{2}As_{2}}$\cite{Harriger11}. 
Starting from an effective overdamped anisotropic $J_{1a}-J_{1b}-J_{2}$ model, they argue that the high-energy cross section cannot be fitted with $J_{1a}=J_{1b}$ and isotropic Landau damping even in the tetragonal PM phase. 
As a result, they suggest that the tetragonal symmetry is broken at high energies and that the peak-splitting is a signature of nematic short range order. 
A similar detailed analysis was made by Ewings \emph{et al.}~on another parent compound $\mathrm{SrFe_{2}As_{2}}$ \cite{Ewings11}.
They find that while an effective $J_{1a}-J_{1b}-J_{2}$ model with fixed parameters is unable to describe the data for different temperatures and energies, a band-structure calculation similar to Park \emph{et al.}\cite{Park10} correctly describes the complete spectrum of magnetic
excitations, without requiring tetragonal symmetry-breaking.

In this paper, we study the spin fluctuation spectrum at high energies in Ba(Fe$_{1-x}$Co$_x$)$_2$As$_2$ with $x=0.047$ using INS. 
In combination with the results in the literature on $x=0$, $0.065$, and $0.074$, the new data allows for a discussion of the systematic evolution of the high energy spin susceptibility as a function of both composition and temperature covering undoped, underdoped, and optimally doped samples.
We find that the transverse splitting is relatively insensitive to composition, suggesting that it is impervious to fine tuning of the Fermi surface and unlikely to arise from static incommensurability \cite{Park10}. 
The splitting is also insensitive to temperature, including temperatures above and below $T_\mathrm{N}$, \textit{i.e.}, it is insensitive to the presence of long-range AFM order. 
This temperature independence was used to suggest that the spin fluctuations are characteristic of the orthorhombic phase at all temperatures, implying a tetragonal symmetry-breaking above $T_N$\cite{Harriger11}.
Here, we adopt the opposite point of view, and suggest that the transverse splitting is a consequence of the anisotropic spin fluctuations already present in the tetragonal-PM phase, manifested by anisotropies in both the magnetic correlation length and Landau damping. 
A similar conclusion was drawn by Refs.~\onlinecite{Ewings11} and \onlinecite{Park11} using first-principle calculations combined with random phase approximation (RPA) and dynamical mean field theory (DMFT) approaches, respectively. 
Here, we support our claims by developing a phenomenological low-energy model for the dynamical magnetic susceptibility derived from an effective three-band model for the Fe sublattice. 
The main advantage of this approach is to provide a simple description of the anisotropic spin fluctuations in terms of a few physically relevant parameters. 
We are able to fit all our INS data using a single set of parameters, showing the suitability of this phenomenological model as a description of magnetic spectrum of the iron pnictides.

Our paper is organized as follows: details of our experiment are contained in Section \ref{sExperiment}; observed spin excitations are described in Section \ref{sObservations}; effects of composition are discussed in Section \ref{sComposition}; the models used to fit the observed spin excitations are detailed in Section \ref{sModels}; details of our analysis of the neutron scattering data are presented in Section \ref{sAnalysis}; finally, our results are discussed in Section \ref{sDiscussion}.

\section{Experimental \label{sExperiment}}

Inelastic neutron scattering measurements were performed on the ARCS\cite{Abernathy12} spectrometer at the Spallation Neutron Source at Oak Ridge National Laboratory on Ba(Fe$_{1-x}$Co$_{x}$)$_{2}$As$_{2}$ with $x=0.047$. 
The sample consists of ten co-aligned single-crystals grown from excess FeAs and CoAs, as outlined in Ref.~\onlinecite{Ni08}, with a total mass of $\sim 2$ grams mounted with a horizontal [$HHL$] scattering plane (in tetragonal notation). 
The sample undergoes a tetragonal to orthorhombic structural transition at $T_\mathrm{S}=60$\,K, orders antiferromagnetically at $T_\mathrm{N}=47$\,K and magnetic long-range order coexists with superconductivity below $T_\mathrm{c}=17$\,K. 
The sample was aligned with the $c$-axis along the incident beam direction and measurements were performed with an incident neutron energy of $E_{i}=$ 250 meV.
The inelastic scattering spectra were measured at $T=5$\,K (\textit{i.e.} in the orthorhombic, AFM and superconducting state) and $T=$ 70 K (\textit{i.e.} in the paramagnetic, tetragonal state). 
Throughout the manuscript, the neutron scattering data are described in the tetragonal $I4/mmm$ coordinate system with 
$\mathbf{Q}=\frac{2\pi}{a}\left(H+K\right)\hat{\imath}+\frac{2\pi}{a}\left(H-K\right)\hat{\jmath}+\frac{2\pi}{c}L\hat{k} = \left(H+K,H-K,L\right)$
where $a=3.95$\,\AA~and $c=12.95$\,\AA. 
In tetragonal $I4/mmm$ notation, $\mathbf{Q}_{\mathrm{AFM}} = \left(\frac{1}{2},\frac{1}{2},1\right)$ [$H$=$\frac{1}{2}$, $K$=$0$]. 
$H$ and $K$ are defined to conveniently describe diagonal cuts in the $I4/mmm$ basal plane as varying $H$ ($K$) corresponds to a longitudinal $\left[H,H\right]$ scan (transverse $\left[K,-K\right]$ scan) through $\mathbf{Q}_{\mathrm{AFM}}$.

\section{Description of the spin excitations in the $x=0.047$ sample \label{sObservations}}

The elastic magnetic scattering and low-energy spin dynamics ($< 15$\,meV) of this same sample were measured by triple-axis neutron scattering techniques and published elsewhere.\cite{Pratt09,Pratt10}
The $x=0.047$ sample has a small ordered moment ($\sim0.2 \mu_{B}$) and the low-energy spin dynamics in the AFM phase at $T=25$\,K (normal state, $T_\mathrm{c}<T<T_\mathrm{N}$) are characterized by strong damping. 
Unlike similar measurements on the \emph{A}Fe$_{2}$As$_{2}$ parent compounds where a sharp spin gap appears in the ordered state, the spin gap is strongly broadened or absent for the $x=0.047$ composition, suggesting that the damping energy scale is comparable to the spin gap energy in the ordered state ($\sim10$\,meV). 
Time-of-flight measurements are required in order to ascertain if well-defined collective spin wave excitations appear in this system at higher energy transfers.

\begin{figure}
\begin{centering}
\includegraphics[width=0.9\linewidth]{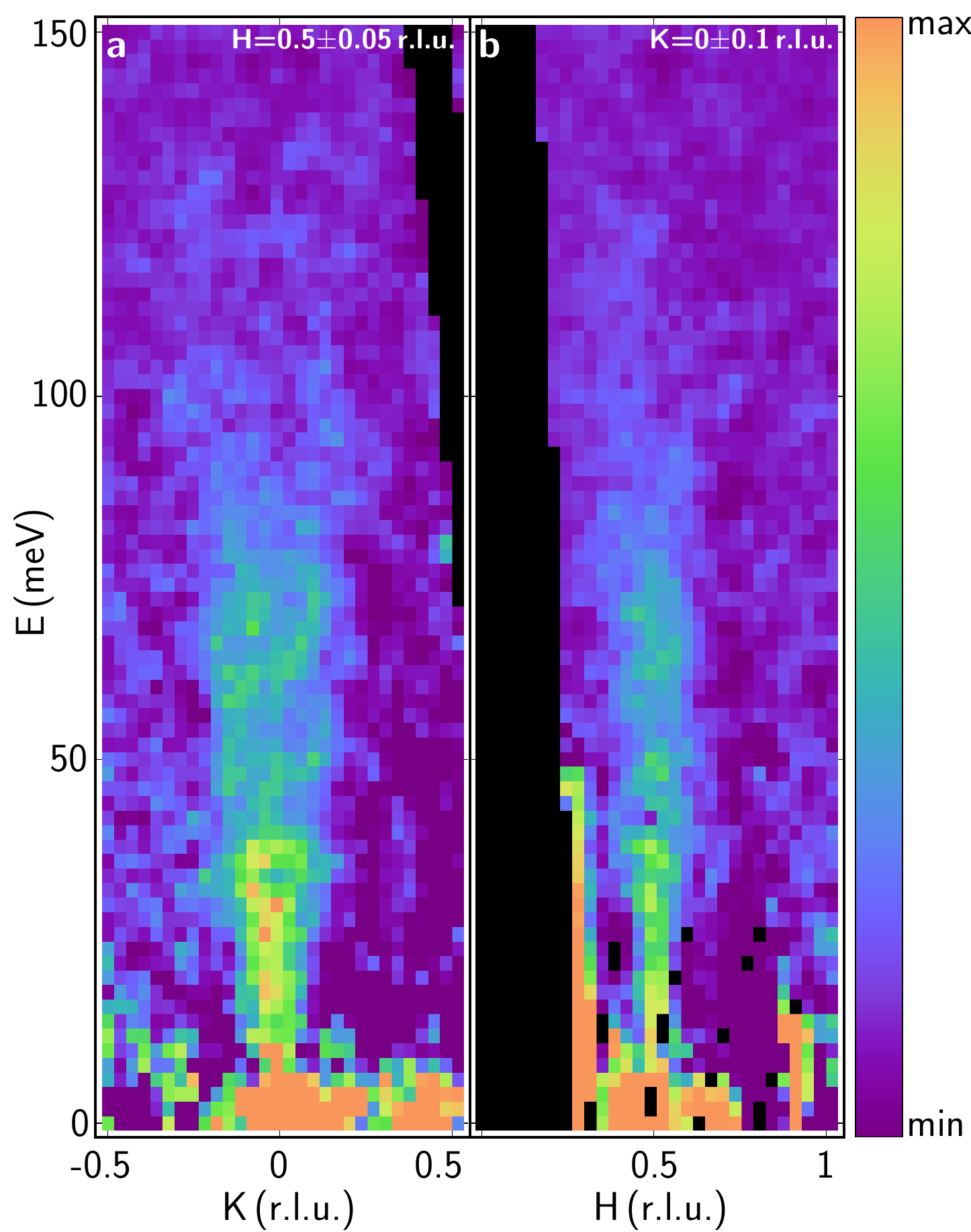} 
\par\end{centering}

\caption{(color online) 
Inelastic neutron scattering data measured on Ba(Fe$_{0.953}$Co$_{0.047}$)$_{2}$As$_{2}$ at $T=5$\,K using ARCS with $E_\mathrm{i} = 250$\,meV and the crystal aligned with the incident neutron beam along the [0,0,1] direction. 
(a) Transverse slice of the data along the $[K,-K]$ direction through $\mathbf{Q}_{\mathrm{AFM}}=(\frac{1}{2},\frac{1}{2},L)$ after averaging over $H=0.5\pm0.05$ r.l.u. 
(b) Longitudinal slice of the data along the $[H,H]$ direction through $\mathbf{Q}_{\mathrm{AFM}}=(\frac{1}{2},\frac{1}{2},L)$ after averaging over $K=0\pm0.1$ r.l.u. 
In each panel, the color scale represents the intensity of scattered neutrons. 
\label{fig1}}
\end{figure}

Time-of-flight measurements of the spin excitations at $T=5$\,K are shown in Fig.~\ref{fig1}.
The magnetic excitations are observed to emanate from $\mathbf{Q}_\mathrm{AFM}$ and are steeply dispersive, extending to energies approaching 150\,meV. 
Surprisingly, we were not able to observe sharp and well defined collective spin wave excitations at $T=5$\,K at any energy despite the long range AFM order. 
This observation confirms that strong damping is a general feature of the spectrum, in agreement with the low-energy triple-axis results.\cite{Pratt09,Pratt10} 
The damping of the spin waves is larger than that found in CaFe$_2$As$_2$\cite{Diallo09,Zhao09} and BaFe$_2$As$_2$\cite{Harriger11} where long-lived collective modes are seen above $\sim 50$\,meV and provide clear evidence of a conical spin wave dispersion.  
In addition a well-defined spin gap is also observed in CaFe$_2$As$_2$ and BaFe$_2$As$_2$. 
In this respect, the spin fluctuation spectrum for $x=0.047$ is more like that measured in the optimally doped and paramagnetic compositions of Ba(Fe$_{1-x}$Co$_x$)$_2$As$_2$ ($x>0.06$), where the spin fluctuations are short-ranged and diffusive in character.\cite{Inosov09,Li10}
\begin{figure}
\begin{centering}
\includegraphics[width=1\linewidth]{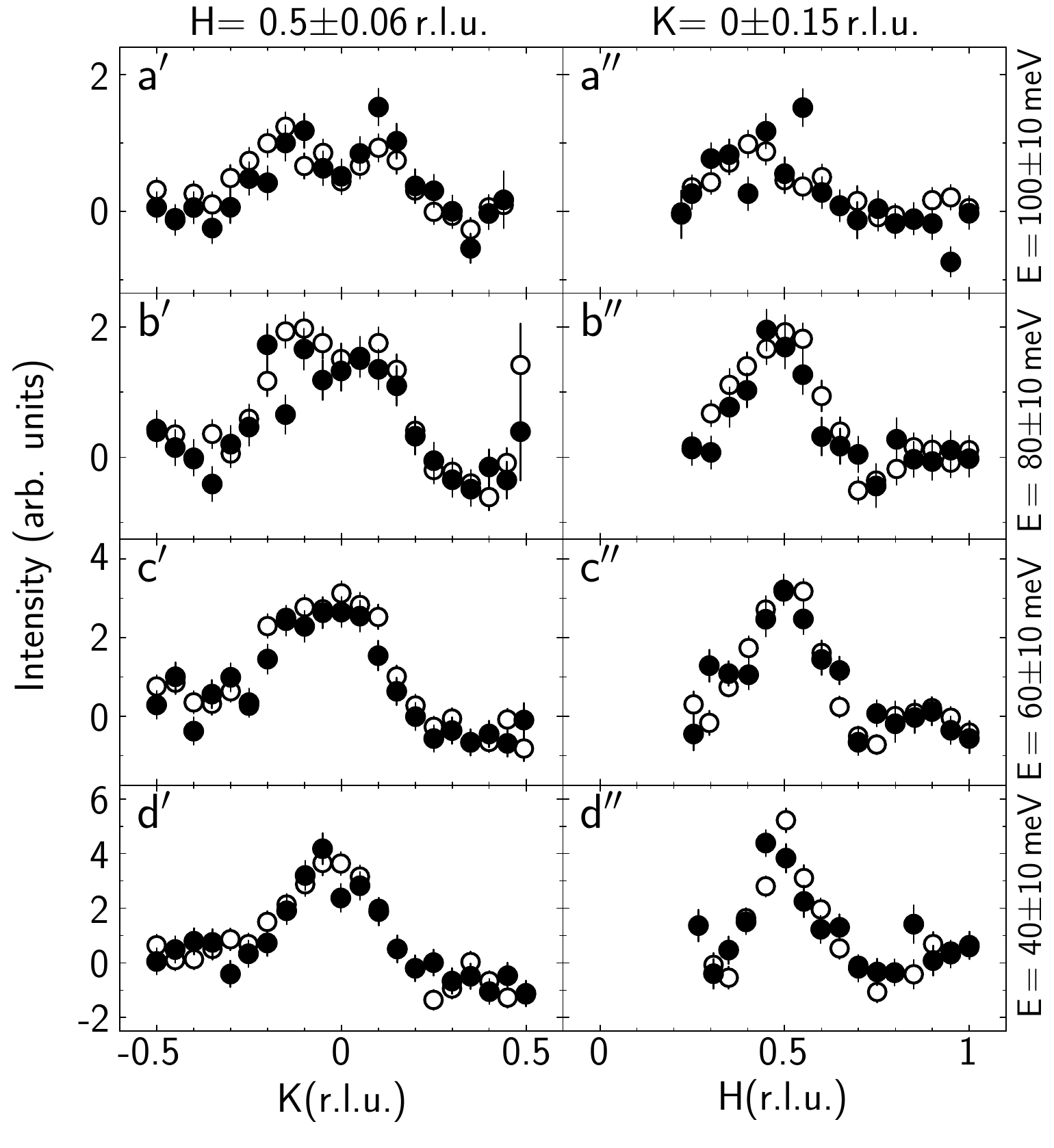} 
\par\end{centering}

\caption{
Comparison of inelastic neutron scattering data measured on Ba(Fe$_{0.953}$Co$_{0.047}$)$_{2}$As$_{2}$ at $T=$ 5 K (empty circles) and $T=$ 70 K (filled circles). 
Transverse cuts of the data at (a$'$) 100 $\pm$ 10 meV, (b$'$) 80 $\pm$ 10 meV, (c$'$) 60 $\pm$ 10 meV, (d$'$) 40 $\pm$ 10 meV. 
Longitudinal cuts of the data at (a$''$) 100 $\pm$ 10 meV, (b$''$) 80 $\pm$ 10 meV, (c$''$) 60 $\pm$ 10 meV, (d$''$) 40 $\pm$ 10 meV. 
\label{fig2}}
\end{figure}

\begin{figure*}
\begin{centering}
\includegraphics[width=0.95\textwidth]{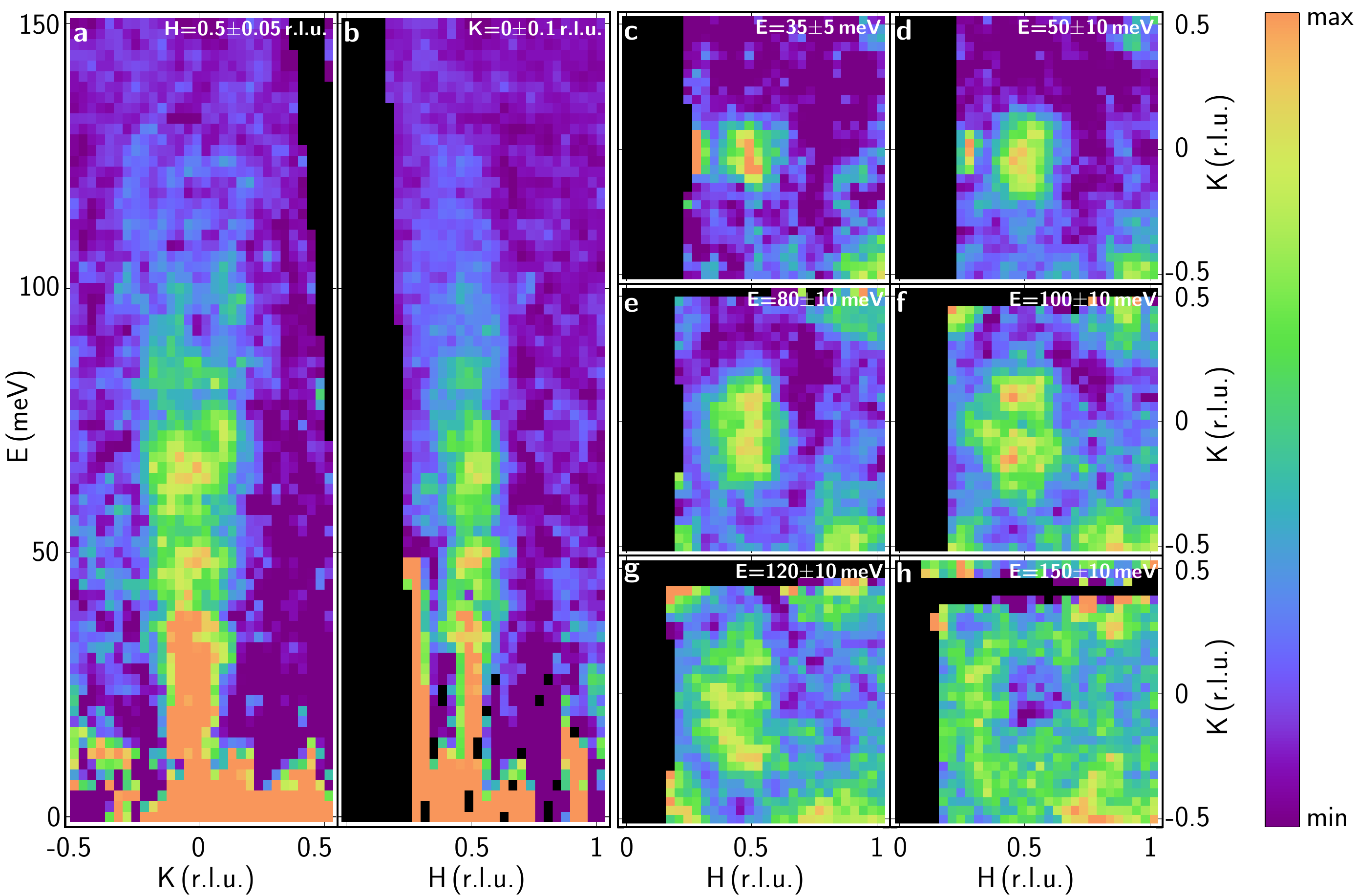} 
\par\end{centering}

\caption{(color online) 
Inelastic neutron scattering data measured on Ba(Fe$_{0.953}$Co$_{0.047}$)$_{2}$As$_{2}$ using ARCS with $E_\mathrm{i} = 250$\,meV and the crystal aligned with the incident neutron beam along the [0,0,1] direction. 
Data at $T=5$\,K and 70\,K are summed together to improve statistics. 
(a) Transverse slice of the data along the $[K,-K]$ direction through $\mathbf{Q}_{\mathrm{AFM}}=(\frac{1}{2},\frac{1}{2},L)$ after averaging over $H=0.5\pm0.05$ r.l.u. 
(b) Longitudinal slice of the data along the $[H,H]$ direction through $\mathbf{Q}_{\mathrm{AFM}}=(\frac{1}{2},\frac{1}{2},L)$ after averaging over $K=0\pm0.1$ r.l.u. 
Constant energy slices in the $(H,K)$-plane averaged over an energy range of (c) $35\pm5$\,meV, (d) $50\pm10$\,meV, (e) $80\pm10$\,meV, (f) $100\pm10$\,meV, (g) $120\pm10$\,meV, and (h) $150\pm10$\,meV. 
In each panel, the color scale represents the intensity of scattered neutrons, where the maximum intensity in each panel, in arb. units, is (a) 10, (b) 10, (c) 10, (d) 8, (e) 5, (f) 3, (g) 2, (h) 1; and the minimum intensity in all panels is -1 arb. units.
\label{fig3}}

\end{figure*}

The diffusive nature of the spin fluctuations for $x=0.047$ in the AFM state is further emphasized by measurements in the paramagnetic state at $T=70$\,K $>T_\mathrm{N}$ ($T_\mathrm{S}$), shown in Fig.~\ref{fig2}, which display a magnetic spectrum nearly identical to the one at $T=5$\,K. 
Thus, we find that despite long-range AFM order, the response function has a diffusive character, with the spin fluctuations being characterized by strong damping and short correlation length as opposed to sharp collective spin-wave excitations. 
In addition, a comparison of the transverse and longitudinal cuts shown in Figs.~\ref{fig2}(a$'$)-(d$'$) and (a$''$)-(d$''$) confirm a substantial anisotropy of the in-plane correlation lengths and a transverse splitting of the high energy spin fluctuations both above and below $T_\mathrm{N}$. 
Similar features were reported previously for the $x=0.074$ composition. 
The similarity of the spin fluctuations at the two temperatures allows us to average the data taken at $T=5$\,K and $70$\,K in order to improve the counting statistics for subsequent analysis described below. Figure \ref{fig3}(a) and (b) show temperature-averaged spectrum cut in the transverse and longitudinal directions, respectively, while panels (c)-(h) show constant energy slices of the data in the $(H,K)$ plane.
In particular, the slices at $80$ (Fig.~\ref{fig3}(e)) and $100$\,meV (Fig.~\ref{fig3}(f)) clearly display the transverse splitting of the spin excitations.

\section{Composition dependence \label{sComposition}}

The high-energy spin fluctuations have been studied in several compositions of the Ba(Fe$_{1-x}$Co$_x$)$_2$ series ($x=0$\cite{Harriger11}, $0.065$\cite{Lester10}, $0.074$\cite{Li10}, $0.08$\cite{Inosov09}, and
the $x=0.047$ composition discussed here) as well as in the parent compound CaFe$_2$As$_2$.\cite{Diallo09,Zhao09,Diallo10} 
In each composition, the spin excitations are seen to extend up to energies $>100-150$\,meV. 
In the paramagnetic phase, the spin fluctuations in momentum space display a pronounced elliptical anisotropy within the Fe layer, with softer magnetic excitations extending along the $[K,-K]$-direction (transverse to $\mathbf{Q}_\mathrm{AFM}$) and stiffer excitations along the $[H,H]$-direction (longitudinal to $\mathbf{Q}_\mathrm{AFM}$). 
In this regime, the anisotropy is defined in terms of the magnetic correlation lengths along antiferromagnetic (longitudinal correlation length, $\xi_\mathrm{L}$) and ferromagnetic bonds (transverse correlation length, $\xi_\mathrm{T}$) of the stripe magnetic structure:

\begin{equation}
\eta_{\xi}=\frac{\xi_\mathrm{L}^{2}-\xi_\mathrm{T}^{2}}{\xi_\mathrm{L}^{2}+\xi_\mathrm{T}^{2}}\label{eta}
\end{equation}

$\eta_{\xi}=0$ corresponds to isotropic spin fluctuations while $\eta_{\xi}=1$ ($\eta_{\xi}=-1$) corresponds to the extreme limit where $\xi_\mathrm{L}\gg\xi_\mathrm{T}$ ($\xi_\mathrm{L}\ll\xi_\mathrm{T}$).
This two-fold anisotropy at $\mathbf{Q}_\mathrm{AFM}=(\frac{1}{2},\frac{1}{2},1)$ is allowed by the simple tetragonal symmetry of the Fe sublattice.\cite{Park10} 
The origin of the anisotropy can be deduced from both localized\cite{Pallab11} and itinerant descriptions of the magnetism in these materials.
In the limit of local-moment magnetism, one can use the $J_{1}-J_{2}$ model to show that $\eta_{\xi}=J_{1}/2J_{2}$.\cite{Diallo10} 
Within this approach, $\eta_{\xi}=0$ corresponds to no coupling between the two interpenetrating N\'eel sublattices making up the stripe AFM structure ($J_{1}=0$), whereas $\eta_\xi=1$ corresponds to the classical stability limit of the stripe ordered antiferromagnetic state, which gives way to G-type magnetic order for $J_{1}>2J_{2}$. 
In the itinerant approach, $\eta_{\xi}$ is a consequence of the ellipticity of the electron pockets. 

For those compositions having long-range AFM order, the spin-waves dispersion is anisotropic within the Fe layer. 
To compare the anisotropy in the AFM and PM states, we define the anisotropy of the spin wave velocity as
\begin{equation}
\eta_{c}=\frac{c_\mathrm{L}^{2}-c_\mathrm{T}^{2}}{c_\mathrm{L}^{2}+c_\mathrm{T}^{2}}
\label{etac}
\end{equation}
where $c_\mathrm{L}$ ($c_\mathrm{T}$) is the spin wave velocity along the longitudinal (transverse) direction, respectively. 
If one uses a local-moment description for the spin wave dispersion, \textit{i.e.}, the $J_{1a}-J_{1b}-J_2$ model, and if the orthorhombic distortion is small ($a_\mathrm{O}\approx b_\mathrm{O}$), then $\eta_{c}$ is given by 
\begin{equation}
\eta_{c}=\frac{J_{1a}+J_{1b}}{4J_{2}+J_{1a}-J_{1b}}
\end{equation}

When $J_{1a}\approx J_{1b}=J_1$, the spin-wave velocity anisotropy reduces to the same result as that obtained for the correlation lengths in the tetragonal paramagnetic phase, $\eta_c=\eta_\xi=J_1/2J_2$.

Using these expressions, we can compute experimental values for the correlation length and spin wave velocity anisotropies for several different iron arsenide compositions in both the ordered and paramagnetic phases, as shown in Table \ref{tbl1}.
In those systems with long-range AFM order, the anisotropy of the low-energy spin fluctuations does not change strongly above $T_\mathrm{N}$, \textit{i.e.}, between the AFM-orthorhombic and PM-tetragonal phases.
This observation casts much doubt on the ``nematic spin fluid'' model proposed in Ref.~\onlinecite{Harriger11}. 
In that model, local orthorhombic distortions are proposed to exist based on the analysis of short-wavelength zone boundary spin-waves at $\mathbf{Q}=(1,0,L)$ using the $J_{1a}-J_{1b}-J_{2}$ model. 
The temperature independence of these zone boundary spin-waves suggests that $J_{1a}$ and $J_{1b}$, which are very different in the orthorhombic phase, remain so even in the tetragonal paramagnetic phase.
However, this local symmetry breaking cannot hold in the long-wavelength limit, where the crystallographic symmetry must be respected and an average nearest-neighbor exchange $2J_{1}=J_{1a}+J_{1b}$ must be assumed. 
Such a scenario would predict a sizable change in the momentum space anisotropy of low energy spin excitations above and below $T_\mathrm{N}$, since 
\begin{equation}
\frac{\eta_{c}^{\mathrm{tet}}}{\eta_{c}^{\mathrm{ort}}}=1+\frac{J_{1a}-J_{1b}}{4J_{2}}
\label{etac_Js}
\end{equation}

Using the values of the exchange constants $J_{1a}$ and $J_{1b}$ extracted from the spin-wave fitting of BaFe$_{2}$As$_{2}$, we have $\eta_{c}^{\mathrm{tet}}/\eta_{c}^{\mathrm{ort}}\approx1.7$. 
Thus, we would expect a change of the in-plane anisotropy by 70\%, however, no change of the anisotropy is observed experimentally.\cite{Harriger11}
This temperature independence suggests that the $J_{1}-J_{2}$ model is an appropriate parameterization of the spin dynamics in the long-wavelength limit (in either the tetragonal or orthorhombic phases). 
At shorter wavelengths/higher energies, the local-moment models clearly run into trouble \cite{Ewings11} and a description that explicitly takes into account the itinerancy of the magnetic moments is needed.\cite{Park11} 
It is interesting to note that spin fluctuations become more anisotropic with Co doping in the Ba122 system, reaching values $\eta_\xi>\frac{1}{2}$ for $x>0.065$.

At energies $> 80 - 100$\,meV, the anisotropic spin fluctuations split along the transverse direction and seem to form separate counter-propagating branches. 
We note that this splitting is unusual and not typical of propagating spin waves, where constant energy contours will form a ring of scattering from collective modes that propagate in all directions. 
This transverse splitting phenomenon was first noted in the $x=0.074$ composition\cite{Li10} and has since been observed in $x=0$ parent compound.\cite{Harriger11}
In the current study on $x=0.047$, we also observe the transverse splitting, as shown in Figures \ref{fig2} and \ref{fig3}. 
We note that the energy where this transverse splitting takes place and its intensity are temperature-independent (Fig.~\ref{fig2}), \textit{i.e.}, the splitting is unmodified even deep in the AFM ordered state, a feature also observed in BaFe$_{2}$As$_{2}$.\cite{Harriger11} 
Viewed as a propagating mode where $c_\mathrm{T}= \hbar\omega/\Delta q$, Table \ref{tbl1} shows that the velocity of this mode is weakly dependent on composition within the Ba122 series, with a value of $316$\,meV\,\AA\ at $x=0$ that softens to $\sim 240$\,meV\,\AA\ for optimally doped superconducting compositions $x>0.065$. 
The weak dependence of this splitting on composition casts some doubt on its interpretation as an incipient incommensurability, {\it i.e.}, a band nesting effect, which should be very sensitive to composition.\cite{Park10}
Recently, static incommensurate AFM order was indeed observed, but only in a very narrow range of Co concentration from $0.056<x<0.06$.\cite{Pratt11}

\begin{table}
\caption{
Composition dependence of the transverse propagation velocity and anisotropy of spin fluctuations in $A$(Fe$_{1-x}$Co$_x$)$_2$As$_2$ ($A=$ Ca, Sr, Ba). 
For undoped compounds in the antiferromagnetic ordered (AF) phase, the transverse velocity and anisotropy parameters are obtained from a fit to a model of propagating spin waves. 
For paramagnetic (P) compounds, the parameters are obtained from either damped spin wave or diffusive models. 
See references for further details.
}

\begin{centering}
\begin{tabular}{|c|c|c|c||c|c|c|}
\hline 
$A$ & $x$  	&$T$ (K)& state & $c_\mathrm{T}$ (meV \AA)  & $\eta$  & Ref. \\
\hline 
Ca  & 0  	& 5		& AFM	& 370 (10)	& 0.32 (2)	& \onlinecite{Diallo09}		\\
Ca  & 0  	& 10	& AFM	& 350 (40)	& 0.34 (10)	& \onlinecite{Zhao09}		\\
Ca  & 0  	& 180	& PM	& -			& 0.28 (18)	& \onlinecite{Diallo10}		\\
Sr  & 0  	& 6		& AFM	& 335 (20)	& 0.21 (4)	& \onlinecite{Ewings11}		\\
Ba  & 0  	& 7		& AFM	& 316 (11)	& 0.41 (2)	& \onlinecite{Harriger11}	\\
Ba  & 0  	& 150	& PM	& $\sim$ 300& $\sim$ 0.4& \onlinecite{Harriger11}	\\
Ba  & 0.047	& 5		& AFM	& 230 (30)	& 0.35 (9)	& {this work}				\\
Ba  & 0.047	& 70	& PM	& $''$		& $''$		& {this work}				\\
Ba  & 0.065	& 7		& PM	& 230 (30)	& 0.7 (2)	& \onlinecite{Lester10}		\\
Ba  & 0.074	& 5		& PM	& 245 (10)	& 0.5 (1)	& \onlinecite{Li10}			\\
Ba  & 0.075	& 200	& PM	& -			& 0.7		& \onlinecite{Park10}		\\
\hline
\end{tabular}\label{tbl1} 
\par\end{centering}
\end{table}

\section{Modeling the spin excitations in the paramagnetic phase \label{sModels}}

\subsection{Diffusive model}

To describe the magnetic spectrum in the paramagnetic state, we first consider an itinerant model that captures the diffusive character of the spin dynamics. 
One can perform a first-principles calculation for the spin dynamics by using the complete density functional theory (DFT) band structure and incorporating the electronic interaction via RPA\cite{Park10,Ewings11} or DMFT.\cite{Park11} 
Indeed, such calculations provide a good description of the magnetic spectrum of different compounds, but their complexity makes it difficult to sort out the essential physics responsible for the behavior of the magnetic spectrum. 

Here, instead of considering the full band structure input, we adopt a phenomenological approach.
The low-energy magnetic spectrum of this model has been discussed in Refs.~\onlinecite{Fernandes10} and \onlinecite{Eremin10}. 
Near $\mathbf{Q}_{\mathrm{AFM}}$ in the $I4/mmm$ tetragonal coordinate system, we obtain the complex dynamical susceptibility:

\begin{equation}
\chi\left(\mathbf{q}+\mathbf{Q}_{\mathrm{AFM}},\omega\right)=\frac{\chi_{0}}{\xi^{-2}+\left(q^{2}+2\eta_{\xi}q_{x}q_{y}\right)-i\omega\gamma_{\mathbf{q}}}
\label{eq1}
\end{equation}

Here, $\xi$ is the bare magnetic correlation length and $\chi_{0}$ is the inverse of the typical magnetic energy scale. 
The imaginary term appears due to the decay of spin excitations (paramagnons) in particle-hole pairs, and $\gamma_{\mathbf{q}}$ is the corresponding momentum-dependent Landau damping. 
The momentum-space anisotropy term $\eta_{\xi}$ is an indirect consequence of the ellipticity of the electron pockets considered in this model. 
Taking the imaginary part, which is the quantity measured by INS, we can cast the result in the form:

\begin{equation}
\chi''\left(\mathbf{q}+\mathbf{Q}_{\mathrm{AFM}},\omega\right)=\frac{\chi_{0}\xi^2\,\omega\Gamma_{\mathbf{q}}}{\omega^{2}+\Gamma_{\mathbf{q}}^{2}(1+\xi_{\mathbf{q}}^{2}q^{2})^{2}}
\label{damping}
\end{equation}
with 
\begin{equation}
\xi_{\mathbf{q}}^{2}q^{2}=\xi^{2}(q^{2}+2\eta_{\xi}q_{x}q_{y})
\label{corlength}
\end{equation}
and 
\begin{equation}
\Gamma_{\mathbf{q}}=\frac{1}{\gamma_{q} \xi^{2}}=\Gamma\left[1+\alpha^{2}\left(q^{2}+2\eta_{\Gamma}q_{x}q_{y}\right)\right]
\label{gammaq}
\end{equation}

In Eqn.~(8), we expand the momentum-dependent damping $\Gamma_{\mathbf{q}}$ for small deviations from $\mathbf{Q}_{\mathrm{AFM}}$, introducing the anisotropy parameter $\eta_{\Gamma}$. 
To ensure that $\Gamma_{\mathbf{q}}^{-1}$ is maximum at $q=0$, we restrict our analysis to $\left|\eta_{\Gamma}\right|<1$. 
Similarly, to ensure that the magnetic correlation length in Eqn.~(\ref{corlength}) is well-defined, we assume $\left|\eta_{\xi}\right|<1$. 
Notice that Eqn.~(\ref{corlength}) naturally gives rise to Eqn.~(\ref{eta}) considered before.

\begin{figure*}
\begin{centering}
\includegraphics[width=0.95\textwidth]{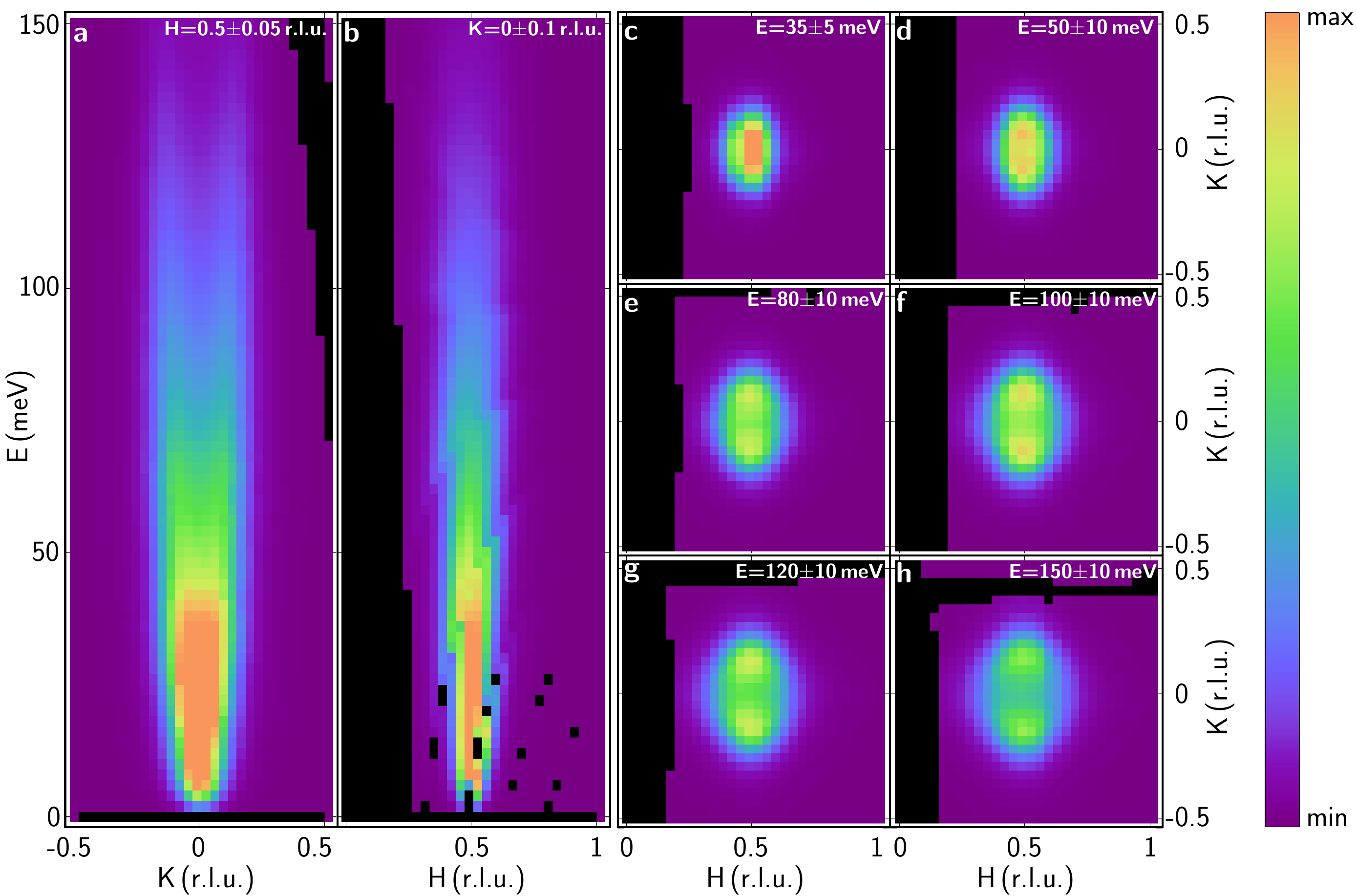} 
\par\end{centering}

\caption{(color online) 
Calculations of the inelastic neutron scattering spectrum as obtained from a model of itinerant diffusive spin fluctuations, as described by Eqn.~(\ref{damping}). 
The parameters of the model were obtained by fits to the neutron scattering data for Ba(Fe$_{0.953}$Co$_{0.047}$)$_{2}$As$_{2}$ as shown in Figs.~\ref{fig6} and \ref{fig7}. 
(a) Transverse slice of the data along the $[K,-K]$ direction through $\mathbf{Q}_{\mathrm{AFM}}=(\frac{1}{2},\frac{1}{2},L)$ after averaging over $H=0.5\pm0.05$ r.l.u. 
(b) Longitudinal slice of the data along the $[H,H]$ direction through $\mathbf{Q}_{\mathrm{AFM}}=(\frac{1}{2},\frac{1}{2},L)$ after averaging over $K=0\pm0.1$ r.l.u. 
Constant energy slices in the $(H,K)$-plane averaged over an energy range of (c) $35\pm5$\,meV, (d) $50\pm10$\,meV, (e) $80\pm10$\,meV, (f) $100\pm10$\,meV, (g) $120\pm10$\,meV, and (h) $150\pm10$\,meV. 
In each panel, the color scale represents the intensity of scattered neutrons, where the maximum intensity in each panel, in arb. units, is (a) 10, (b) 10, (c) 10, (d) 8, (e) 5, (f) 3, (g) 2, (h) 1; and the minimum intensity in all panels is 0.
\label{fig4}}
\end{figure*}

At low energies, this form of the susceptibility correctly captures the elliptical shape of $\chi''\left(\mathbf{q},\omega\right)$ peaked at $\mathbf{Q}=\mathbf{Q}_{\mathrm{AFM}}(\mathbf{q}=0)$, in agreement with the INS measurements on a variety of 122 compounds.\cite{Diallo10,Li10} 
The sign of the ellipticity depends on $\eta_{\xi}$; for the parent and Co-doped $\mathrm{BaFe_{2}As_{2}}$ compounds, such as the one shown in Fig.~\ref{fig3}, the longitudinal correlation length [parallel to $\mathbf{Q}_{\mathrm{AFM}}=(\frac{1}{2},\frac{1}{2})$] $\xi_\mathrm{L}=\xi\sqrt{1+\eta_{\xi}}$ is longer than the transverse correlation length [perpendicular to $\mathbf{Q}_{\mathrm{AFM}}=(\frac{1}{2},\frac{1}{2})$] $\xi_\mathrm{T}=\xi\sqrt{1-\eta_{\xi}}$, implying that $\eta_{\xi}>0$.
On the other hand, the sign of $\eta_{\xi}$ depends on the band structure (see, for instance Ref.~\onlinecite{Park10}), and there is {\it a priori} no reason for it to be positive or negative. 
This is to be contrasted to the classical AFM $J_{1}-J_{2}$ model, which always predicts $\eta\propto J_{1}/J_{2}>0$.

At high energies, as $\omega$ increases, the momentum-dependence
of the Landau damping eventually leads to the splitting of the single
peak of $\chi''\left(\mathbf{q},\omega\right)$. Along the transverse
direction, the low-energy peak splits into two symmetric peaks for
energies $\omega>\omega_\mathrm{T}$, with:

\begin{equation}
\omega_\mathrm{T}=\Gamma\sqrt{1+\frac{2\xi^{2}\left(1-\eta_{\xi}\right)}{\alpha^{2}\left(1-\eta_{\Gamma}\right)}}\label{omega_LO}\end{equation}

Similarly, along the longitudinal direction, the peak-splitting takes
place for energies above

\begin{equation}
\omega_\mathrm{L}=\Gamma\sqrt{1+\frac{2\xi^{2}\left(1+\eta_{\xi}\right)}{\alpha^{2}\left(1+\eta_{\Gamma}\right)}}.\end{equation}

Therefore, if $\eta_{\xi}=\eta_{\Gamma}$, the elliptical shape of $\chi''\left(\mathbf{q},\omega\right)$ at low energies evolves in an elliptical-ring structure for $\omega>\omega_\mathrm{T}=\omega_\mathrm{L}$.
On the other hand, for $\eta_{\xi}>\eta_{\Gamma}$ ($\eta_{\xi}<\eta_{\Gamma}$) the splitting happens at a lower energy along the transverse (longitudinal) direction. 
Thus, depending on the magnitude of the difference $\Delta\omega\equiv\omega_\mathrm{T}-\omega_\mathrm{L}$, there can be a wide regime of energies where the low-energy single-peak response of $\chi''\left(\mathbf{q},\omega\right)$ at $\mathbf{Q}_{\mathrm{AFM}}$ splits into two peaks equidistant from $\mathbf{Q}_{\mathrm{AFM}}$. 
For $\Delta\omega>0$ (\textit{i.e.} $\eta_{\xi}>\eta_{\Gamma}$), these two peaks split along the direction transverse to $\mathbf{\mathbf{Q}_{\mathrm{AFM}}}$, whereas for $\Delta\omega<0$ (\textit{i.e.} $\eta_{\xi}<\eta_{\Gamma}$), the two peaks split along the longitudinal direction to $\mathbf{\mathbf{Q}_{\mathrm{AFM}}}$.
Therefore, the experimental data in the parent and Co-doped $\mathrm{BaFe_{2}As_{2}}$ samples (Fig.~\ref{fig3}) are compatible with $\eta_{\xi}>\eta_{\Gamma}$.

Figure \ref{fig4} shows calculations of the neutron scattering cross-section for the diffusive model using parameters determined from fits of the neutron data for the $x=0.047$ compound to Eqn.~(\ref{damping}).
The fits to the data are described in detail below. 
This figure shows the same slices of the neutron intensity as in Fig.~\ref{fig3} and can be compared directly to the data. 
In general, we find that the neutron data is best described with anisotropy parameters $\eta_{\xi}>0$ and $\eta_{\Gamma}<0$, resulting in anisotropic ellipsoids of scattering at low energies and a transverse splitting at higher energies. 

\begin{figure*}
\begin{centering}
\includegraphics[width=0.95\textwidth]{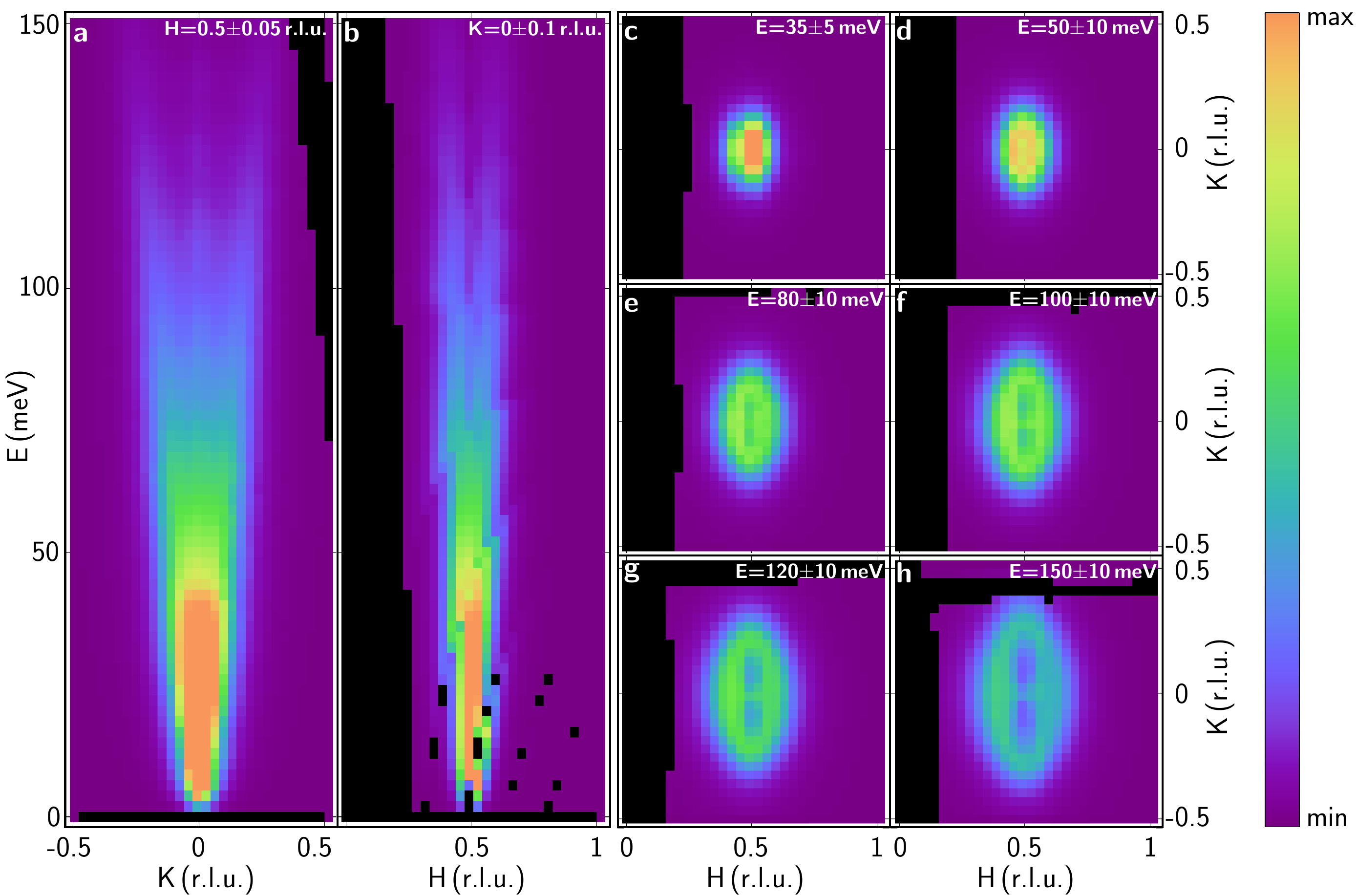} 
\par\end{centering}

\caption{(color online) 
Calculations of the inelastic neutron scattering spectrum as obtained from a ballistic model of overdamped spin waves, as described by Eqn.~(\ref{ballistic}). 
The parameters of the model were obtained by fits to the neutron scattering data for Ba(Fe$_{0.953}$Co$_{0.047}$)$_{2}$As$_{2}$ as shown in Figs.~\ref{fig6} and \ref{fig7}. 
(a) Transverse slice of the data along the $[K,-K]$ direction through $\mathbf{Q}_{\mathrm{AFM}}=(\frac{1}{2},\frac{1}{2},L)$ after averaging over $H=0.5\pm0.05$ r.l.u. 
(b) Longitudinal slice of the data along the $[H,H]$ direction through $\mathbf{Q}_{\mathrm{AFM}}=(\frac{1}{2},\frac{1}{2},L)$ after averaging over $K=0\pm0.1$ r.l.u. 
Constant energy slices in the $(H,K)$-plane averaged over an energy range of (c) $35\pm5$\,meV, (d) $50\pm10$\,meV, (e) $80\pm10$\,meV, (f) $100\pm10$\,meV, (g) $120\pm10$\,meV, and (h) $150\pm10$\,meV. 
In each panel, the color scale represents the intensity of scattered neutrons, where the maximum intensity in each panel, in arb. units, is (a) 10, (b) 10, (c) 10, (d) 8, (e) 5, (f) 3, (g) 2, (h) 1; and the minimum intensity in all panels is 0.
\label{fig5}}

\end{figure*}

\subsection{Ballistic model with propagating modes}

Besides the purely diffusive dynamics discussed in the previous section, one can consider a different phenomenological form for the spin susceptibility.
In the so-called ballistic model, the dynamics is governed by propagating overdamped spin-waves in the disordered paramagnetic state 
\begin{equation}
\chi''\left(\mathbf{q}+\mathbf{Q}_\mathrm{AFM},\omega\right)=\frac{\chi_{0}\xi^2\,\omega\Gamma}{\omega^{2}+\Gamma^{2}(1+\xi_{\mathbf{q}}^{2}q^{2}-\frac{\xi_{\mathbf{q}}^{2}}{c_{\mathbf{q}}^{2}}\omega^{2})^{2}}
\label{ballistic}
\end{equation}
with anisotropic spin-wave velocity:

\begin{equation}
c_{\mathbf{q}}^{2}q^{2}=c^{2}(q^{2}+2\eta_{c}q_{x}q_{y})
\end{equation}

The anisotropy in the velocity, $\eta_{c}$, is the same as that introduced in Eqn.~(\ref{etac}). 
Presumably, the spin wave velocities in the paramagnetic phase are comparable to those found deep in the AFM ordered state.
The form of the susceptibility in Eqn.~(\ref{ballistic}) is diffusive in nature (consisting of relaxational dynamics and a single peak response) at low energies, where the spin-wave wavelength is longer than the correlation length (\textit{i.e.} $\omega<c_{\mathbf{q}}\xi^{-1}$). 
The renormalized spin-wave modes will appear in the form of a broad elliptical ring of scattering at constant $\omega$ when $\omega>c_{\mathbf{q}}\xi^{-1}$ is satisfied.
The high-energy form of the scattering then represents a section of the damped, conical spin-wave dispersion. 
This approach was used to describe the anisotropic quasi-propagating mode postulated in Ref.~\onlinecite{Li10}.
A similar approach was also used in Ref.~\onlinecite{Harriger11}, although in that case a local moment Heisenberg model was employed to describe the spin wave dispersion whereas here we assume a linear dispersion.

Figure \ref{fig5} shows calculations of the neutron scattering cross-section for the ballistic model using parameters determined from fits of the neutron data for the $x=0.047$ compound to Eqn.~(\ref{ballistic}). 
The fits to the data are described in detail below. 
This figure shows the same slices of the neutron intensity as the data in Fig.~\ref{fig3}, as well as the  diffusive model in Fig.~\ref{fig4}. 
We see that the anisotropy of the spin wave velocity with $\eta_{c}>0$ describes the elliptical shape of the scattering cross-section. 
At high energies, the scattering assumes the expected form of an elliptical ring. 
We note that while the spectra in panels (a) and (b) of Figs.~\ref{fig4}--\ref{fig5} appear similar, there is a qualitative difference in the higher energy slices in panels (c)-(h). 
The diffusive model correctly reproduces the transverse splitting observed by experiment, whereas the ballistic model can only produce elliptical rings or, in the extreme case of $\eta_{c}\approx1$, an elliptical ring that is pinched in the middle (see Fig.~\ref{fig5}(g) and (h)).

\section{Analysis of the neutron scattering data \label{sAnalysis}}

We now describe the procedure used to fit the INS data to both the diffusive and ballistic models.  The inelastic neutron scattering data is directly related to the imaginary
part of the magnetic susceptibility through the relation

\begin{equation}
S(\mathbf{Q},\omega)\propto f^{2}(Q)\chi''(\mathbf{Q},\omega)(1-e^{-\hbar\omega/kT})^{-1}
\label{eq_S}
\end{equation}

Each scattered neutron collected at ARCS is recorded as an individual
event, where the time-of-flight and detector position is stored.
It is impossible to visualize the data without first converting its
format from a list of flight times and pixel positions to a spectrum
for each detector pixel, this necessitates collecting the data into energy bins such that each spectrum is a histogram of the number
of detected neutrons per unit energy transfer. This process is often
referred to as data reduction and was in this case accomplished with
the DANSE software package.\cite{danse}

The observed inelastic neutron scattering intensity $I(\mathbf{Q},\omega)$ includes contributions from various non-sample sources -- primarily the crystal holder and the sample environment. 
We also consider non-magnetic scattering from the sample itself as a source of background.
The collective intensity from these sources comprise the background function $B(\mathbf{Q},\omega)$.
The background function was estimated by averaging intensity from equal $Q$ and $\omega$ sections of the detector which were far from magnetic intensity.
\begin{equation}
B(Q_{i},\omega_{i})=\frac{\sum_{j}B(Q_{j},\omega_{j})\delta(Q_{i}-Q_{j})\delta(\omega_{i}-\omega_{j})}{\sum_{j}\delta(Q_{i}-Q_{j})\delta(\omega_{i}-\omega_{j})}
\end{equation}
where the sum goes over all detectors far from magnetic intensity and the $Q_{i,j},\omega_{i,j}$ correspond to bin centers. 
This approximate to $B(\mathbf{Q},\omega)$ can only properly account for contributions to the background which are isotropic about the incident beam direction and is, more accurately, $B(Q,\omega)$. 
An estimate of the sample-scattered intensity can be obtained by subtracting the averaged background estimate from the observed intensity.
\begin{equation}
S(\mathbf{Q},\omega)=I(\mathbf{Q},\omega)-B(Q,\omega)
\end{equation}

The background subtracted spectra for detectors within the range $0.25<H<0.75$\,r.l.u., $-0.5<K<0.45$\,r.l.u.~and $20<E<150$\,meV were collectively fit with the Levenberg-Marquardt algorithm to each model function, Eqns.~(\ref{damping}) and (\ref{ballistic}). 
In addition to either of the model functions, a background was fit to account for an observed residual component of $B(\mathbf{Q},\omega)$ caused by a detector bank top-to-bottom asymmetry of the measured intensity, the fit residual background function was 
\begin{equation}
B_{r}(\mathbf{Q},\omega)=-a_{0}e^{-\hbar\omega/a_{1}}K
\end{equation}
where the $a_{i}$ are positive model parameters and $K$ (in r.l.u.) is a measure of the vertical displacement from the horizontal detector plane.
Our fitting routine ensured that, for all cases, $\left|\eta\right| \le 1$.
Volumetric $H$,$K$,$E$ data was fit in order to avoid artifacts introduced by binning which are inherent in 1-D cuts and 2-D slices. 
Fit parameter values and their associated errors, plus values of $\eta_\xi$ derived from the fit parameters are presented in Table \ref{table_fitres}. 
We observed that fits of the data were rather insensitive to changes in $\eta_\Gamma$, and subsequently fixed $\eta_{\Gamma}$ to $-1$ for our fitting. 
This limit gives $q$-independent (constant) damping in the longitudinal direction ($\Gamma$) and $q$-dependent damping in the transverse direction ($\Gamma_q=\Gamma(1+2\alpha^2q^2)$).

\begin{table}[h]
\begin{centering}
\begin{tabular}{|c|c||c|c|}
\hline 
\multicolumn{2}{|c||}{Diffusive (Eqn.~\ref{damping})}  & \multicolumn{2}{|c|}{Ballistic (Eqn.~\ref{ballistic})} \\
\hline 
$\chi_{0}\xi^2$\,(arb. units)	& $5.0(4)$		& $\chi_{0}\xi^2$\,(arb. units)	& $6.3(7)$ 	\\
$\Gamma$\,(meV)				& $10.7(11)$	& $\Gamma$\,(meV)			& $8.5(11)$ 	\\
$\xi_\mathrm{L}$\,(\AA)		& $9.4(5)$		& $\xi_\mathrm{L}$\,(\AA)	& $11.8(9)$ 	\\
$\xi_\mathrm{T}$\,(\AA)		& $5.0(3)$		& $\xi_\mathrm{T}$\,(\AA)	& $8.2(6)$ 	\\
$\alpha$\,(\AA)				& $2.0(2)$		& $c$\,(meV\,\AA)			& $450(40)$ 	\\
$\eta_\Gamma$				& $-1$			& $\eta_{c}$  				& $0.73(6)$ \\
\hline 
$\eta_{\xi}$				& $0.57(6)$		& $\eta_{\xi}$  			& $0.35(9)$ \\
\hline
\end{tabular}
\caption{Fit parameter values and calculated $\eta_\xi$ for the two models: diffusive (Eqn.~\ref{damping}), and ballistic (Eqn.~\ref{ballistic}).}
\label{table_fitres} 
\end{centering}
\end{table}

\begin{figure}
\begin{centering}
\includegraphics[width=0.95\columnwidth]{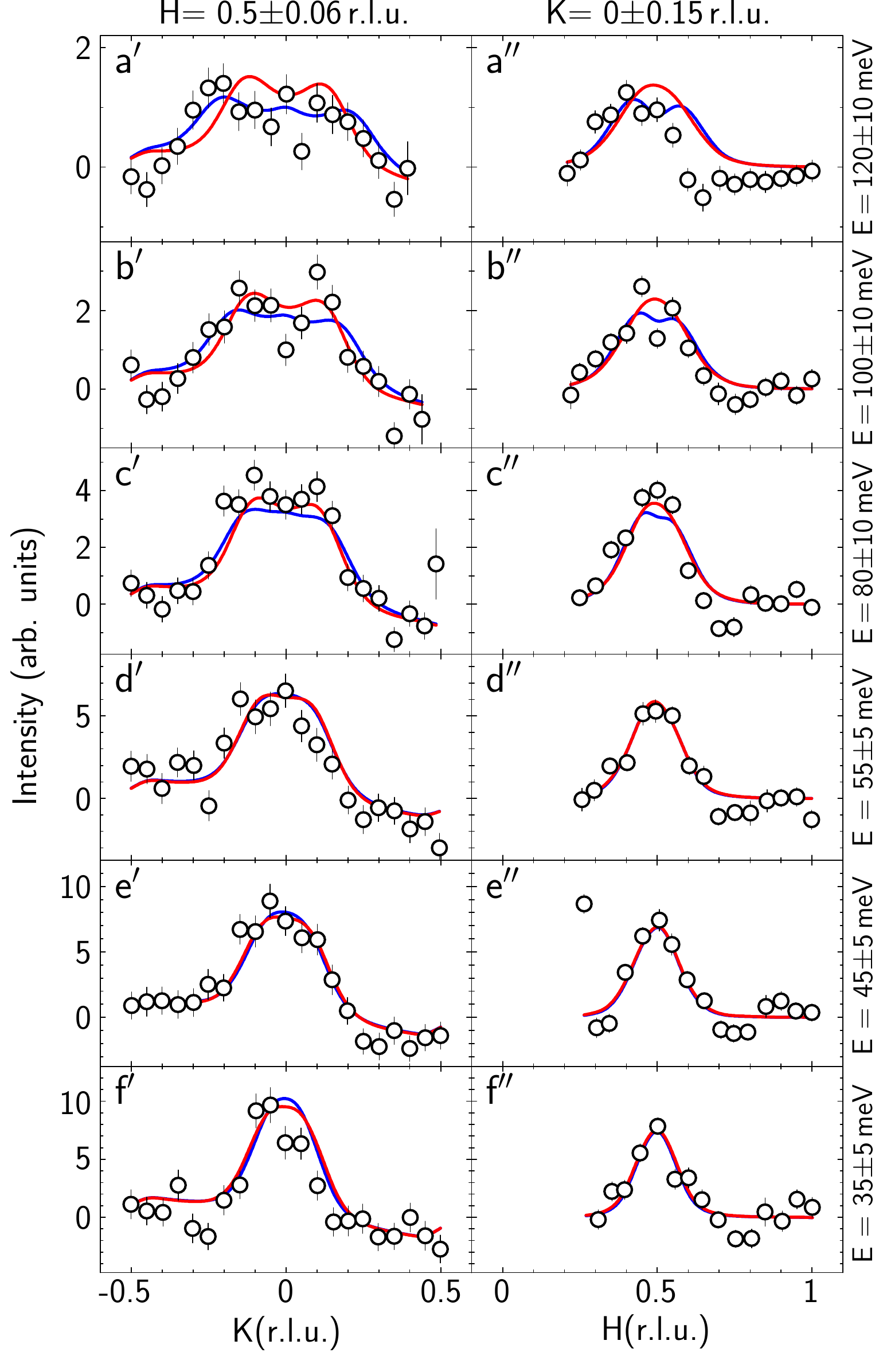} 
\par\end{centering}
\caption{
(color online) Fitting results for the diffusive (red lines) and ballistic (blue lines) models to the summed $T=5$\,K and $70$\,K data (open symbols). 
Transverse cuts ($'$) and longitudinal cuts ($''$) through $\mathbf{Q}=(\frac{1}{2},\frac{1}{2},L)$ at (a) $E=120\pm10$\,meV, (b) $E=100\pm10$\,meV, (c) $E=80\pm10$\,meV, (d) $E=55\pm5$\,meV, (e) $E=45\pm5$\,meV, and (f) $E=35\pm5$\,meV.
\label{fig6}}
\end{figure}

\begin{figure}
\begin{centering}
\includegraphics[width=0.95\columnwidth]{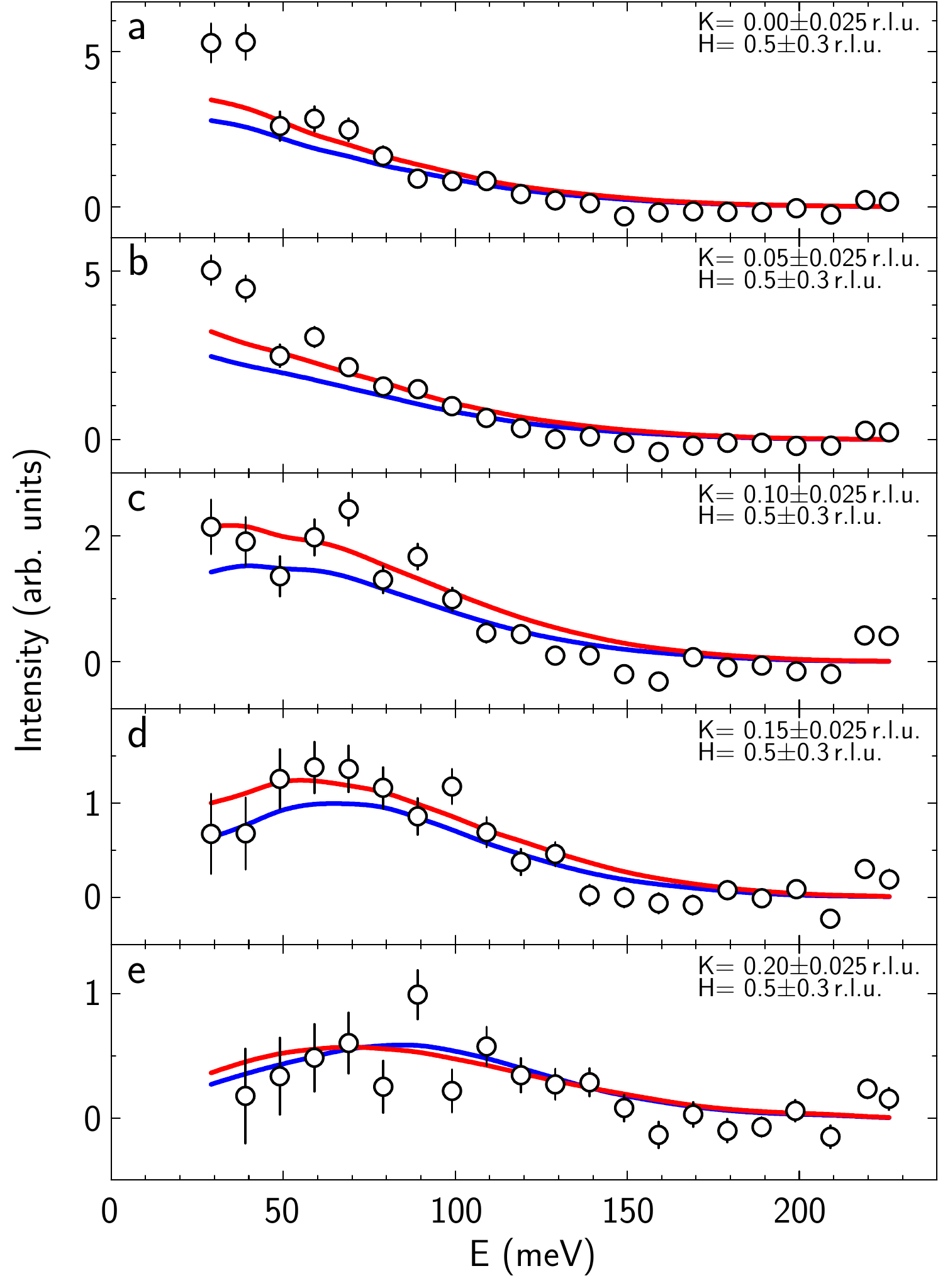} 
\par\end{centering}
\caption{(color online) Fitting results for the diffusive (red lines) and ballistic (blue lines) models to the summed $T=5$\,K and $70$\,K data (open symbols). 
Spectra cuts centered at (a) $(0.5,0.5,L)$, (b) $(0.55,0.45,L)$, (c) $(0.6,0.4,L)$, (d) $(0.65,0.35,L)$, and (e) $(0.7,0.3,L)$.
\label{fig7}}
\end{figure}
This manuscript describes inelastic neutron scattering data from Ba(Fe$_{0.953}$Co$_{0.047}$)$_2$As$_2$ which exhibit a transverse splitting in the spin excitations which has been previously observed for other doping concentrations of Co. 
A detailed understanding of the spin fluctuations is an important component of understanding the origin of superconductivity in the doped BaFe$_2$As$_2$ compounds.
We find that this high energy splitting is present in the presence of both a paramagnetic-tetragonal and antiferromagnetic-orthorhombic-superconducting phase.
Furthermore, we have successfully described our data by means of a model involving anisotropic Landau damping and magnetic correlation length, which does not break the high temperature tetragonal symmetry of the system.
Our findings show that it is not necessary to introduce symmetry breaking, quasi-propagating modes, or incommensurability in order to explain the high energy splitting of the spin excitations in Ba(Fe$_{1-x}$Co$_x$)$_2$As$_2$, which readers of PRB should find important and interesting.

We use Eqn.~(\ref{eq_S}) for the neutron intensity to test both the diffusive and the ballistic models, Eqns.~(\ref{damping}) and (\ref{ballistic}) respectively. 
The fittings are presented in Figs.~\ref{fig6} and \ref{fig7} (red curve for the diffusive and blue curve for the ballistic model) for fixed energy and fixed momentum, respectively. 
For energies below $60$\,meV, the two models give nearly identical line shapes, which is not surprising, since both models reduce to an identical description of diffusive spin excitations at low energies.
On the other hand, at higher energies, the purely diffusive model is in better agreement with the data, since the longitudinal cut has a maximum at $q=0$ whereas the transverse cut has a minimum at $q=0$. 
The ballistic model, on the other hand, gives the opposite: a minimum at $q=0$ for the longitudinal cut and a maximum at $q=0$ for the transverse cut.

\section{Discussion \label{sDiscussion}}

According to the fits of the INS data, neither the purely diffusive nor the ballistic model is strongly favored over the other. 
Mostly, this is due to the broad and overdamped nature of the spin excitations and the convergence of the two models at low energies. 
Surprisingly, even though both models are low-energy descriptions of the magnetic excitation spectrum, the fitting to the data reveals that they are able to capture some of the intricate physics responsible for the high-energy behavior. 
Physically, the two models represent two different conceptual pictures for AFM fluctuations in the paramagnetic phase.
The ballistic model would represent spin wave propagation in disordered AFM, whereas the diffusive model including Landau damping is based more closely on a quasiparticle description obtainable from either the simplified 3-band structure discussed here, or by first principles calculations in Refs.~\onlinecite{Park10} and \onlinecite{Ewings11}. 
However, the qualitative form of the scattering throughout the $(H,K)$-plane, in particular the observation of the split transverse modes at high energies, seems to favor a description of the excitations in terms of the diffusive model, without the need to introduce propagating modes (compare Figs.~\ref{fig2}, \ref{fig3} and \ref{fig4}). 
In the diffusive model, the peak splitting is caused by the interplay between the momentum-space anisotropies associated with the Landau damping and the magnetic correlation length, both of which follow naturally from the band structure of the iron pnictides. 

It is interesting to note that a similar discussion regarding the existence or not of propagating modes took place also in the context of a much simpler material, namely, bcc iron 
(see, for instance, Refs.~\onlinecite{Lynn75} and \onlinecite{Wicksted84}). 
In that case, claims for the existence of propagating modes at low energies followed from observations of splitting in the constant-energy cuts of the INS data similar to our case. 
It was subsequently shown that the nature of the damping in an itinerant ferromagnet, where $\Gamma_{q}\rightarrow0$ as $q\rightarrow0$, can result in a splitting that resembles a spin-wave dispersion.\cite{Wicksted84}
In the ferromagnetic case, energy cuts at constant-momentum should show an inelastic peak in the ballistic model, whereas no such peak will be observed in the diffusive model. 
In our specific case of paramagnetism in the stripe AFM ordered phase, one can see (Fig.~\ref{fig7}) that both the ballistic and purely diffusive models produce an inelastic peak in the constant-momentum energy scan. 
Thus, unlike the ferromagnetic case, the presence or absence of an inelastic peak in the spectrum cannot discern between the two models.

The excitations in the paramagnetic phase of the parent BaFe$_2$As$_2$ compound have been described using a $J_1-J_2$ Heisenberg model combined with a phenomenological momentum-dependent damping function.
This approach was able to describe the spin fluctuations close to the zone boundary only after allowing for tetragonal symmetry-breaking, as discussed in Ref.~\onlinecite{Harriger11}. 
As described above, this ``nematic spin fluid'' model is called into question since the proposed model should lead to large changes in the low energy anisotropy of the spin fluctuations below $T_\mathrm{N}$, which are not observed. 
In order to describe the peculiar transverse splitting, the authors chose a strongly anisotropic form for the Landau damping that resulted in large damping along the longitudinal direction and small damping in the transverse direction. 
This choice effectively washes out the longitudinal spin waves, leading to the split transverse modes. 
In our notation, the authors chose $\eta_{\Gamma}<-1$ meaning that the damping function has a saddle point around $\mathbf{Q}_{\mathrm{AFM}}$, rather than a minimum, as one would expect. 
While similar in spirit to the approach considered here, our model does not require local symmetry breaking in the paramagnetic state. 
We note, however, that our analysis does not preclude the existence of a nematic phase in the iron pnictides. 
To probe the nematic phase, it would be more appropriate to perform INS measurements in detwinned samples (see, for instance \onlinecite{Fernandes12} and also \onlinecite{Si08}).

Finally, we comment on the role of incommensuration of the SDW ordering vector. 
In Ref.~\onlinecite{Park10}, first principles calculations suggest that an incommensurability is present for the parent compound $\mathrm{BaFe_2As_2}$.
It was then proposed that the high-energy peak splitting would be nothing but a manifestation of this incommensurability. 
If indeed the magnetism in the iron pnictides is of itinerant nature, then it is reasonable to expect, on general grounds, the development of incommensurability at some critical doping (see the seminal work of Rice\cite{Rice70}).
Subsequent to those predictions, static incommensurability was indeed observed in a narrow range of doping concentrations in the Co-doped Ba122 system.\cite{Pratt11}
The incommensurability is observed to develop in a first-order fashion with doping concentration.\cite{Pratt11}
However, in the INS data presented here for the underdoped Ba(Fe$_{0.953}$Co$_{0.047}$)$_{2}$As$_{2}$, as well as in the previously presented data for the parent{\cite{Harriger11} and optimally doped Ba(Fe$_{0.926}$Co$_{0.074}$)$_{2}$As$_{2}$\cite{Li10}, the high-energy splitting of the inelastic peaks occurs at roughly the same energy $\omega_\mathrm{split}\approx80$\,meV and momentum in both
the commensurate AFM ordered state and the PM state. 
Thus, it unlikely that the transverse splitting is associated with any instability towards incommensurate order.

In summary, we developed a simple expression able to capture the main properties of the spectrum of the magnetic excitations in these materials.
In particular, we showed that the observed transverse splitting of the inelastic peaks of $\chi''\left(\mathbf{q},\omega\right)$ at high energies, as well as their dispersions, can be attributed to the interplay between the anisotropies of the magnetic correlation length and of the Landau damping. 
This explanation follows naturally from the model band structure we used, and does not require breaking of the tetragonal symmetry, or the introduction of quasi-propagating modes or incommensurability. 
We demonstrated the applicability of this model to the 122 iron pnictides by fitting low and high energy INS data on the underdoped compound Ba(Fe$_{0.953}$Co$_{0.047}$)$_{2}$As$_{2}$.

\begin{acknowledgments}
This research is supported by the U.S. Department of Energy, Office of Basic Energy Sciences, Division of Materials Sciences and Engineering.
Ames Laboratory is operated for the U.S. Department of Energy by Iowa State University under Contract No.~DE-AC02-07CH11358. 
Work at Oak Ridge National Laboratory is supported by U.S. Department of Energy, Office of Basic Energy Sciences, Scientific User Facilities Division.
R.M.F.~acknowledges the valuable support from the NSF Partnerships for International Research and Education (PIRE) program OISE-0968226
\end{acknowledgments}

\end{document}